\documentclass[aps,prl,twocolumn,superscriptaddress]{revtex4-1}

\usepackage[utf8]{inputenc}
\usepackage{amsmath}
\usepackage{graphicx}
\usepackage{color}
\usepackage{xspace}
\usepackage{float}
\usepackage{braket}
\usepackage[normalem]{ulem}

\newcommand{\Li}{\ensuremath{^{6}\text{Li}}\xspace}
\newcommand{\aB}{\ensuremath{a_{\text B}}\xspace}
\newcommand{\G}{\ensuremath{\text G}\xspace}

\begin{document}

\title{Microscopic Observation of Pauli Blocking in Degenerate Fermionic Lattice Gases}
\author{Ahmed Omran}
\email{ahmed.omran@mpq.mpg.de}
\author{Martin Boll}
\author{Timon Hilker}
\author{Katharina Kleinlein}
\author{Guillaume Salomon}
\affiliation{Max-Planck-Institut f\"ur Quantenoptik, 85748 Garching, Germany}
\author{Immanuel Bloch}
\affiliation{Max-Planck-Institut f\"ur Quantenoptik, 85748 Garching, Germany}
\affiliation{Ludwig-Maximilians-Universit\"at, Fakult\"at f\"ur Physik, 80799 M\"unchen, Germany}
\author{Christian Gross}
\affiliation{Max-Planck-Institut f\"ur Quantenoptik, 85748 Garching, Germany}

\begin{abstract}
The Pauli exclusion principle is one of the most fundamental manifestations of
quantum statistics. Here, we report on its local observation in a
spin-polarized degenerate gas of fermions in an optical lattice. We probe the
gas with single-site resolution using a new generation quantum gas microscope
avoiding the common problem of light induced losses. In the band insulating
regime, we measure a strong local suppression of particle number fluctuations
and a low local entropy per atom. Our work opens a new avenue for studying
quantum correlations in fermionic quantum matter both in and out of
equilibrium.
\end{abstract}

\maketitle

Quantum statistics distinguishes between two fundamentally different kinds of
particles: bosons, which condense into a single quantum state at zero
temperature, and fermions, for which multiple occupancy of a single state is
forbidden. As a result, identical fermions seem to repel each other, described
 by an effective Fermi pressure on a macroscopic level~\cite{Truscott2001}.
Microscopically, the Pauli blockade manifests itself in a strong suppression of
density fluctuations~\cite{Mueller2010,Sanner2010} and in antibunching of
density-density
correlations~\cite{Oliver1999,Henny1999,Kiesel2002,Ianuzzi2006,Rom2006,Jeltes2007}.
Specifically, fermions in periodic potentials form a band insulating state with
suppressed number fluctuations on each site when the chemical potential lies in
the band gap. Measuring these number fluctuations can therefore be regarded as a direct
probe of Pauli blocking.

Local fluctuations in periodic potentials have been directly studied with
ultracold bosonic atoms in optical
lattices~\cite{Gemelke2009,Bakr2010,Sherson2010}. Some of these experiments
featured site-resolved fluorescence detection with single atom
sensitivity~\cite{Nelson2007,Bakr2009}, which has proven to be a powerful
method for probing quantum many-body systems. However, such quantum gas
microscopy requires a specialized experimental setup with considerably increased
technical complexity. In particular, the often used fermionic alkali atoms are
difficult to laser cool, making the single-site and atom resolved detection
even more challenging. First results have recently been reported on the
imaging of single fermions in dilute thermal
clouds~\cite{Parsons2015,Haller2015,Cheuk2015,Edge2015}. However, microscopy of
quantum degenerate fermions has so far remained out of reach.

Here, we report on the site-resolved characterization of a spin-polarized
degenerate Fermi gas in an optical lattice. In the band-insulating region, we
measure a strong suppression of local atom number fluctuations, more than one
order of magnitude below the Poisson limit expected for uncorrelated particles.
Based on the measurement of the local occupation statistics, we reconstruct the
spatial entropy distribution in the inhomogeneous samples. We obtained these
results with a conceptionally novel quantum gas microscope based on an additional,
dedicated optical lattice for detection. This provides high flexibility for
future experiments and offers an alternative approach~\cite{Preiss2015a,
Fukuhara2015} to overcome the limitations due to parity
detection~\cite{Bakr2009,Sherson2010}.

Our experiments started in a standard magneto-optical trap of \Li loaded from a
Zeeman slower (for details see supplemental material). We further cooled the
atoms using narrow-line laser cooling on the $323\,$nm line~\cite{Duarte2011,
Sebastian2014}. The resulting $50/50$ spin mixture of fermions in the $\ket{F,
m_F} = \ket{1/2,\pm1/2}$ states was then loaded into an optical trap and
evaporatively cooled at a magnetic field controlled scattering length of
$a_\mathrm{evap}=-290\,\aB$, where $\aB$ is the Bohr radius. Next, we
transported the atoms into a glass cell, where we used a further, vertically
propagating optical dipole trap to locally enhance the density for a second
efficient optical evaporation at $a_\mathrm{evap}$. To reduce the vertical
extension of the cloud, the atoms were transferred into a strongly elliptical
dimple trap with a vertical waist of $w_z = 1.7\,\mu{\rm m}$. Two beams were
shone in from the side, which interfered under a small angle to produce a
vertical optical lattice of $3\,\mu{\rm m}$ spacing as illustrated in Fig.~1(a).
The fermions loaded from the dimple trap populated mostly a single plane of this
optical lattice.  However, for high fidelity microscopy, almost all atoms in
the adjacent planes need to be removed. This was achieved by transferring the
atoms in undesired planes in a vertical magnetic field gradient from the
$\ket{1/2,\pm1/2}$ to the $\ket{3/2, \mp1/2}$ states, which are subsequently
lost due to spin changing collisions. The atoms in the final single
plane were evaporatively cooled at $a_\mathrm{evap}$, using a horizontal
magnetic field gradient. The radial confinement counteracting this gradient was
set independently by the depth of the vertical dipole trap. To produce a
spin-polarized sample, we ramped the magnetic field to $27\,\G$, where the
magnetic moment of only the $\ket{1/2,-1/2}$ spin state vanishes. Then, we
reduced the in-plane confinement until the gradient field completely removed
the $\ket{1/2,+1/2}$ spin component, which we verified using Stern-Gerlach
separation.

\begin{figure}[t] \centering
\includegraphics[width=0.4\textwidth]{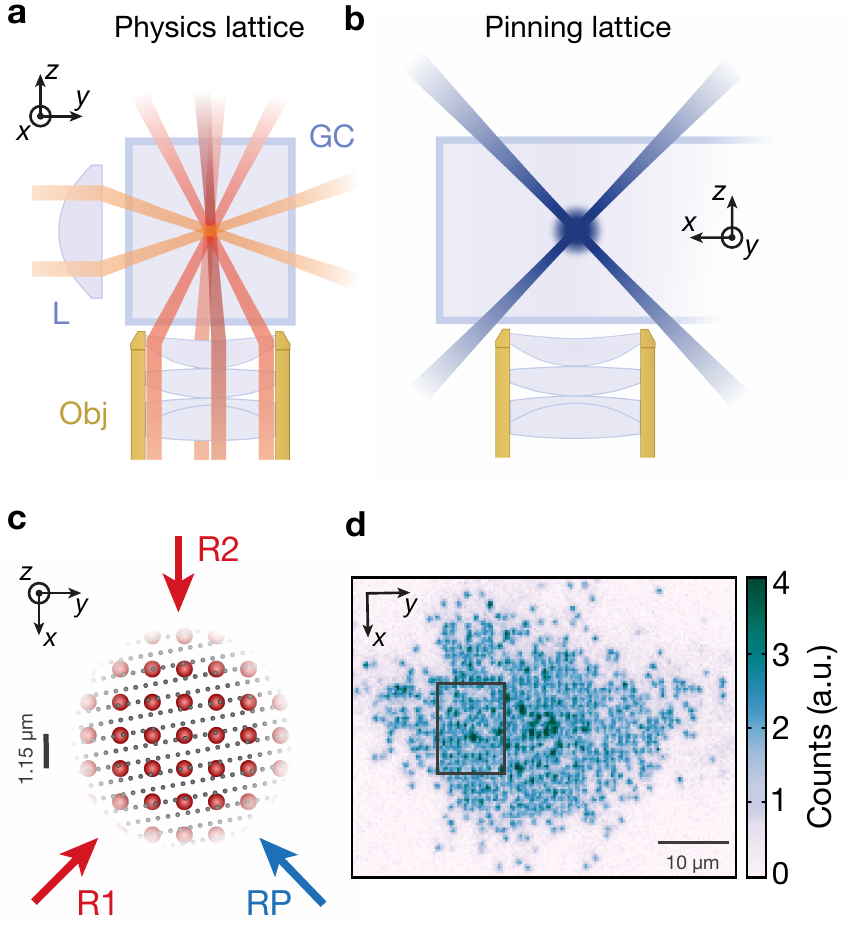}
  \caption{\textbf{Experimental setup.} \textbf{(a)} Geometry of the physics
  lattice. A vertical standing wave of $3\,\mu$m spacing is produced by
  focusing two beams with an aspheric lens (L) onto the atoms from the side.
  The horizontal lattice with a spacing of $1.15\,\mu$m is formed by two pairs
  of beams focused through the imaging objective (Obj) beneath the glass cell (GC).
  \textbf{(b)} Pinning lattice beams viewed from the side. Two beams propagate
  close to the imaging objective in the \textit{x-z} plane, while the third beam is in
  the \textit{y} direction. All beams are retroreflected, forming a $532\,$nm lattice.
  \textbf{(c)} Pinning lattice oversampling and Raman beam geometry viewed from
  top. For imaging, the atoms are loaded from the physics lattice (red dots) to
  the pinning lattice (gray dots). The Raman beams (red, R1 and R2) and repump beam (blue, RP)
  propagate in the \textit{x-y} plane. For the pinning lattice, we show the sites within
  $\pm250\,$nm from one \textit{x-y} plane of the physics lattice. In the experiment the
  relative position between pinning and physics lattice sites is not controlled.
  \textbf{(d)} Fluorescence image of $N=763(10)$ \Li atoms acquired over $1\,$s
  of Raman sideband cooling. The black box marks the region where we evaluated
  the fluorescence statistics shown in Fig. 2.}
\end{figure}

The optical lattice was tailored to the properties of \Li, especially to its
light mass $m$. In a lattice, the single particle energy scale is given by the
recoil energy $E_R=h^2 / 8 m a_l^2$, such that the lattice constant $a_l$ can
be large when the mass is small. For lithium, fast tunneling time scales in the
kilohertz regime are realized even for $a_l > 1\,\mu$m. Thus, the strongly
correlated regime is experimentally accessible, given the availability of
Feshbach resonances to tune the interactions~\cite{Chin2010}.  We loaded the
atoms into such a large scale two-dimensional ``physics'' lattice with
$a_l=1.15\,\mu$m, produced by focusing phase coherent pairs of parallel beams
with a custom microscope objective (numerical aperture $0.5$, effective focal
length $28\,{\rm mm}$). The same objective was later used to collect the
fluorescence signals during imaging [see Fig.~1(a)].\\ We used a deep additional
``pinning'' lattice with a lattice spacing of $532\,$nm that oversampled the
physics lattice [see Figs.~1(b),~1(c)]~\cite{Shotter2011}. The $1064\,$nm pinning
lattice beams were retroreflected with waists of $56\,\mu{\rm m}$ and an
average power of $22\,{\rm W}$ per direction, resulting in an overall trap
depth of $2.5\,$mK and on-site trap frequencies of $1.3\,$MHz on all axes. Prior
to imaging, we transferred the atoms to the pinning lattice, which provides
sufficient local confinement for Raman sideband
cooling~\cite{Monroe1995,Hamann1998,Kerman2000} with a Lamb-Dicke parameter
$\eta\approx0.23$. The Raman cooling (see supplemental material) was optimized
for high fidelity imaging. The scattered photons provided the fluorescence
signal [see Fig.~1(d)], which we detected with an electron-multiplied CCD
camera~\cite{Patil2014,Lester2014}. Over a detection period of $1\,$s, we
collected on average $350$ photons per atom, corresponding to a fluorescence
rate of about $7\,$kHz. Parasitic background light was negligible during the
$1\,$s exposure.

\begin{figure}[b]
\includegraphics[width=0.44\textwidth]{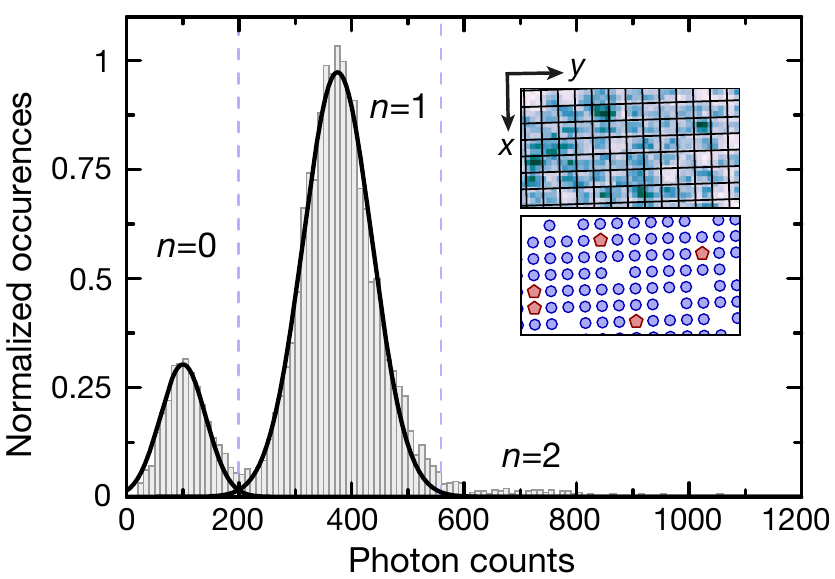}
  \caption{\textbf{Lattice reconstruction.}
  Statistics of fluorescence levels obtained from the amplitude of local
  Gaussian fits per site. On $50$ images, we evaluated the region marked in
  black in Fig.~1d with an average filling of $85\%$. The peaks in the
  histogram of $n=0$ and $n=1$ atoms are clearly separated (black lines are
  Gaussian fits). The few counts to the right are due to double occupancies
  (see supplemental information). The dashed lines mark the thresholds for
  identifying zero, one and two atoms per site. Inset: Close up of an exemplary
  fluorescence image with the results of the reconstruction shown below. Single
  atoms are marked with blue dots, red pentagons indicate double occupancies.}
\end{figure}

\begin{figure*}[t]
\includegraphics[width=0.85\textwidth]{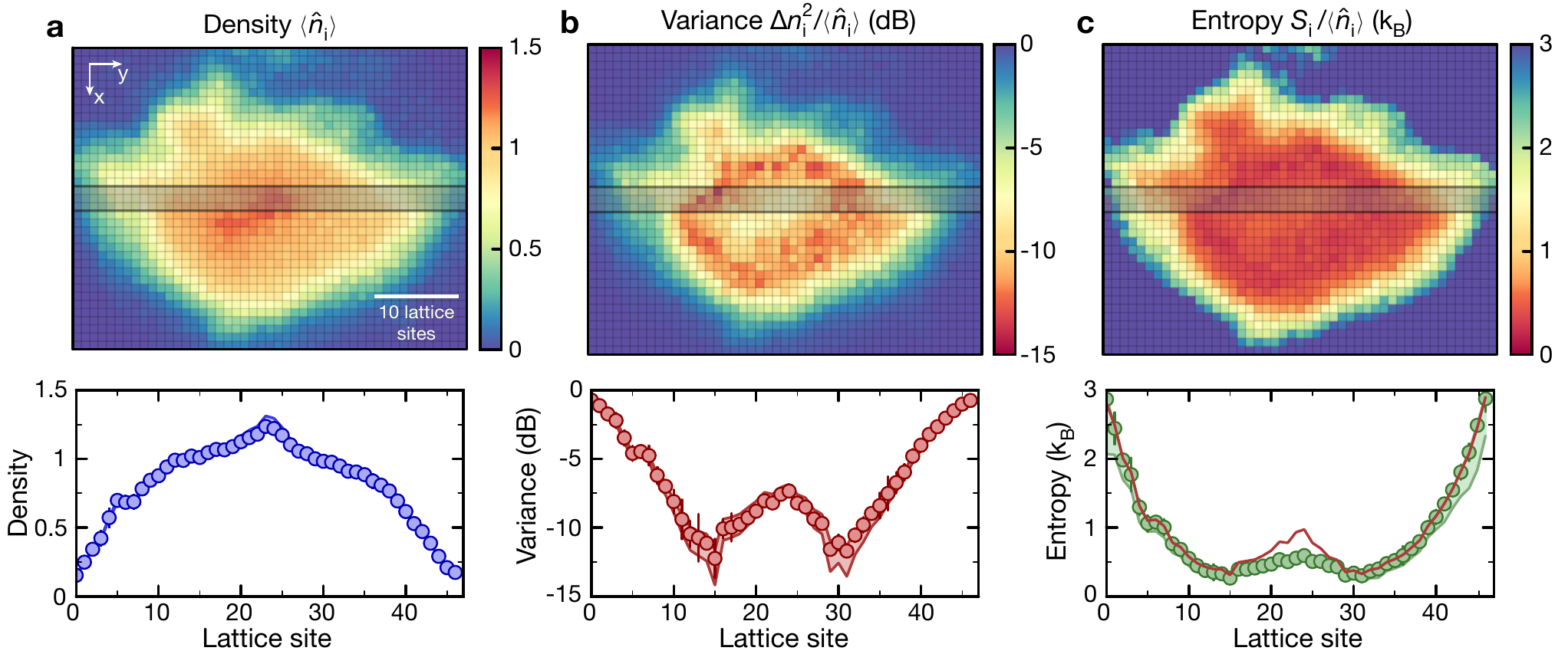}
  \caption{\textbf{Local statistical analysis.} We analyze the local,
  single-site resolved occupation statistics of $425$ images. Each pixel
  corresponds to a physics lattice site in the 2D histograms shown in the top
  row. Bottom row: horizontal cuts averaged over the vertical three sites
  indicated by the transparent box above. The shaded regions correspond to the
  systematic uncertainty in the local population assignment during
  reconstruction. Statistical error bars represent one standard deviation.
  \textbf{(a)} Average density. Pauli blocking results in a large region of
  approximately uniform density around the center.  \textbf{(b)} Normalized
  atom number fluctuations.  The band insulator manifests itself in the
  suppression of fluctuations most strongly in the region of approximately flat
  density in (a).  In the wings of the cloud, Poissonian statistics is
  recovered.  \textbf{(c)} Entropy per atom. The lowest entropies of
  $0.3\,k_{\rm B}$ are found in a ring around the center. The entropy increases
  towards the center and the edge due to the increase of double occupancies and
  holes, respectively. The red line is the result for the entropy, when
  assuming four instead of two accessible single particle states per site.}
\end{figure*}

We reconstruct the physics lattice population via a local image analysis based
on the point-spread function (PSF) of the detection system. We extract the PSF
from fluorescence images of sparsely filled atomic samples by averaging many
isolated single atom signals. This experimentally determined PSF is well
approximated by a Gaussian with a standard deviation of $380(10)\,$nm,
considerably larger than the specified and independently confirmed $290\,{\rm
nm}$. The first step of the reconstruction algorithm is to identify the position
of the physics lattice on each single image by Fourier transforming the image
and extracting the phase associated with the lattice wave vectors. The resulting
physics lattice grid is shown in the inset of Fig.~2. Even though the atoms
were held in the incommensurate pinning lattice during imaging, a clear
distinction of empty and singly occupied physics lattice sites is visible on
the bare fluorescence image. Next, we determine the photon counts in each
physics lattice site from a fit with the Gaussian approximated PSF. The
resulting histogram of the fit amplitudes for images around unity filling [cf.
Fig.~1(d)] per site is shown in Fig.~2. The peaks corresponding to empty and
singly occupied sites are clearly distinct, resulting in a high fidelity of
$99\%$ for discriminating them. Furthermore, the histogram shows several events of
high counts to the right of the $n=1$ peak. We attribute these to doubly
occupied physics lattice sites, which we identify with a reduced fidelity of
$\simeq70\,\%$ (see supplemental material). We expect to improve on the latter
in future experiments, as the detection fidelity for double occupancies is at present
mainly limited by our broadened PSF.\\ 
To estimate the influence of loss and hopping events, we took subsequent images
of the same sparsely populated sample with $1\,$s exposure per image. We found
a loss probability of $2.5(5)\%$ and a tunneling probability of $5(1)\%$
between two images averaged over a large detection region of
$50\times50\,\mu$m, where most of these events occur towards the edge of this
region.

Next, we used the microscope to study the local statistics of spin-polarized,
degenerate fermionic samples. After evaporation and spin polarization, we
ramped up the physics lattice adiabatically within $100\,$ms to $8\,E_{\rm R}$.
Before imaging, we rapidly increased the physics lattice depth to $20\,E_{\rm
R}$ to freeze the atomic distribution. Then we switched on the pinning lattice
linearly within $5\,$ms. Setting the atom number between $N=700$ and
$N=800$ fermions, we obtained images with high filling factors in the center of
the trap [cf. Fig.~1(d)]. We took $425$ images following this protocol and
analyzed the local statistics of population for each lattice site. In the
analysis, we took systematic errors due to the reconstruction as well as
statistical uncertainties into account (see supplemental material).

The fermionic character of the gas is directly visible in the mean density
$\langle \hat{n}_i \rangle$ per site $i$, shown in Fig.~3(a). In an
inhomogeneously trapped sample, Pauli blocking leads to a plateau of unity
filling. The overall shape of the cloud was determined by the trapping
potential, which was not perfectly harmonic due to residual large-scale
imperfections in our physics lattice beams. Even though the spin-polarized
degenerate Fermi gas does not thermalize any more during lattice ramp up, our
loading sequence resulted in a state close to a band insulator. Absence of
thermalization hinders the redistribution of population from the first excited
into the lowest band, which most likely limits the flatness of the observed
density plateau. The atoms in the excited band accumulate in the trap center,
where we detected an increased number of doubly occupied sites.

The incompressibility of the band insulator manifests itself locally as a
suppression of on-site atom number fluctuations $\Delta n^2_i = \langle
\hat{n}_i^2 \rangle - \langle \hat{n}_i \rangle^2$ below the Poissonian
variance given by $\langle \hat{n}_i \rangle$, expected for
uncorrelated atoms. Indeed, in the region of the density plateau we found
strongly suppressed fluctuations with a normalized variance $\Delta
n^2_i/\langle \hat{n}_i \rangle = -11.9_{-2.7}^{+1.6}\,$dB [see Fig.~3(b)],
while the Poisson limit was recovered in the low density wings.

Enabled by the site resolved measurement of the lattice occupations and under
reasonable assumptions on the accessible local quantum states, we reconstructed
the local entropy $S_{i}$, without relying on a local density approximation or
thermal equilibrium.  The entropy per site $S_{i}/k_{\rm B}= - \sum_E p_E^{(i)}
\mathrm{ln}\,p_E^{(i)}$ was calculated from the occupation probabilities
$p_E^{(i)}$ of all many-body energy eigenstates allowed for a given number of
identical fermions distributed over $k$ single particle states. Based on our
observations, we limited the analysis to maximally two atoms per site. For
non-interacting fermions, the probabilities $p_E$ are directly related to the
measured probabilities of finding zero, one or two atoms per site (see
supplemental material). Here, we ignored the finite tunneling, which couples
the different single particle states, leading to an overestimation of the
inferred entropy. The number of populated single particle states was not
directly accessible. Therefore, we analyzed the simplest case $k=2$, and a
second, maximal entropy scenario of $k=4$, assuming the local ground state and
three equally populated excited states, one per spatial direction. In Fig.~3(c),
we present the resulting entropy distribution normalized to the mean site
occupation. The two cases $k=2$ and $k=4$ converge in the limit of low double
occupancy, showing that the entropy can be faithfully reconstructed in this
regime. We extracted a minimum entropy per atom of $S_i/\langle n_i \rangle =
0.3(1)\,k_{\rm B}$ in the density plateau region, while it increased towards
the edge and the center, where the probability to find holes or two atoms was
enhanced, respectively.\\ An upper limit to the temperature in the
two-dimensional harmonic trap $T_{\rm ini}$ prior to lattice ramp-up can be
deduced from the entropy measurement in the lattice. Based on the
experimentally determined total entropy $\sum_i S_i = 1.05(5)\,N k_{\rm B}$, we
obtained $T_{\rm ini}/T_{\rm F} = 0.16(1)$, where $T_{\rm F}$ is the Fermi
temperature~\cite{Carr2004, Castin2007}.

\begin{figure}[b]
\includegraphics[width=0.45\textwidth]{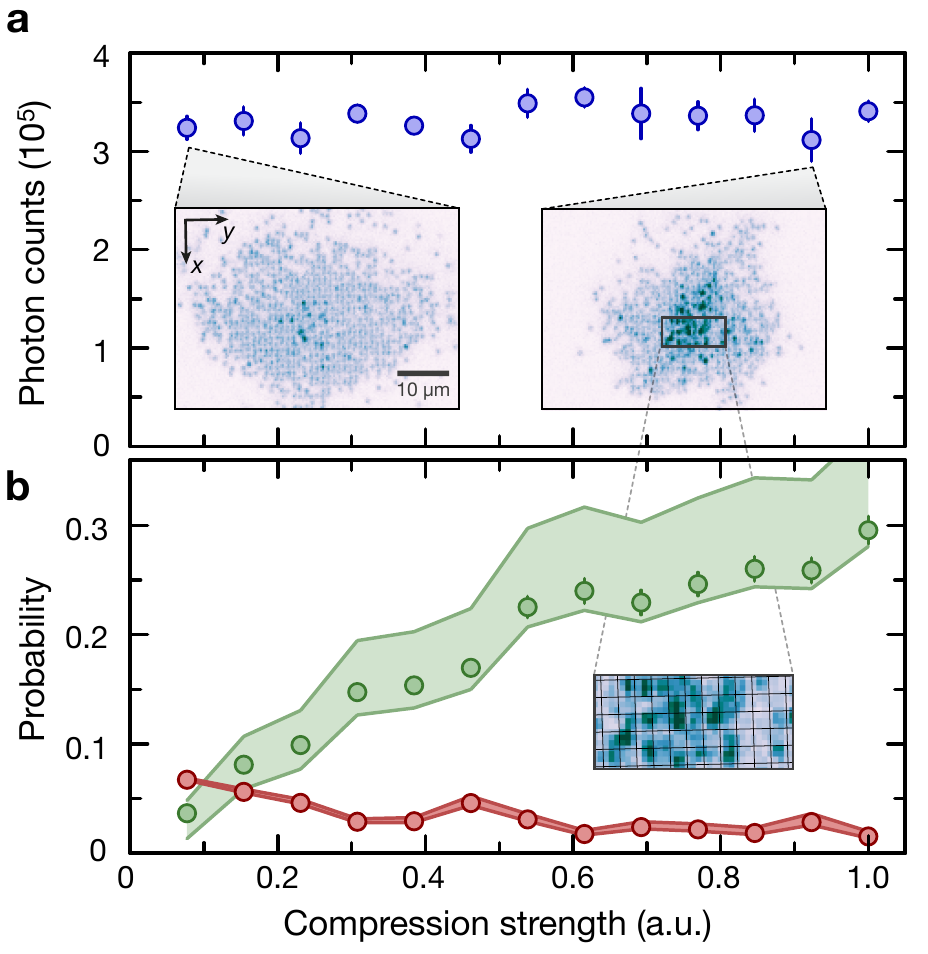}
\caption{\textbf{Absence of parity projection.}
  \textbf{(a)} Overall fluorescence level. We observe a constant total
  fluorescence level while increasing the density by trap compression. Insets:
  Representative images for two different compression strengths. Approximately
  unity filling is observed for weak compression, whereas for stronger
  confinement a dense core of multiply occupied sites is visible. \textbf{(b)}
  Probability of empty (red) and doubly (green) occupied sites versus
  compression strength. The number of holes decreases despite an increase in
  the number of doubly occupied sites. The shaded area corresponds to the
  systematic uncertainty in the local population assignment during
  reconstruction. Inset: Close-up of the central region marked by the black box
  in (a), demonstrating the localization of high fluorescence regions to
  individual lattice sites. Error bars are one standard deviation of the mean.}
\end{figure}

Our measurements described before reveal above unity filling in the center of
the clouds. Contrary to earlier experiments~\cite{DePue1999, Bakr2009,
Sherson2010}, this observation indicates a suppression of parity projection,
that is, the rapid loss of pairs of atoms in the same lattice site during
fluorescence imaging. To set an upper bound on the probability for parity
projection, we increased the density of the spin-polarized sample at constant
atom number and measured the change in fluorescence and local site occupations.
The density was controlled by the confinement in the initial harmonic trap, in
which the gas was still in thermal equilibrium at chemical potential $\mu$.
When ramping up the optical lattice, the atomic sample cannot thermalize.
Consequently, population above the energy $E_R$, around which the band gap
opens, was transferred into the first excited band. Hence, the filling factor
in the center of the lattice increased with increasing $\mu$. We indeed
observed a compression of the cloud and a strong increase of the fluorescence
level in the center of the cloud, but no measurable change in the total
fluorescence signal [see Fig.~4(a)]. This provides strong evidence for a small
probability of parity projection and that the photon scattering rate is constant,
even when adjacent pinning lattice sites were filled -- an important
prerequisite for the detection of higher site occupations. In Fig.~4(b) we
compare the fraction $x_2$ of doubly occupied to the fraction $x_0$ of empty
sites. For the strongest compression, their values reach
$x_2=0.3^{+0.1}_{-0.05}$ and $x_0 = 0.020(5)$, resulting in an upper limit to
parity projection of $9\%$.\\ This observed suppression of parity projection is
explained by the loading from the physics into the pinning lattice. Because of the
shorter lattice constant of the pinning lattice, the number of states per
Brillouin zone is increased compared to the physics lattice. When switching on
the much deeper pinning lattice, states in the first excited band of the
physics lattice connect to states in the lowest band of the pinning lattice.
Hence, initial double occupancies are separated into different pinning lattice
sites as long as the switch on of the pinning lattice is adiabatic, which we
ensured in our experiments.

In conclusion, we presented a site-resolved statistical study of single
component degenerate fermionic lattice gases, directly demonstrating Pauli
blocking in a textbooklike experiment. Additionally, we introduced a novel
quantum gas microscope for fermionic \Li, which separates the detection system
from the physical system under study. Not only does it provide a way to
circumvent parity projection, but it is also directly applicable to more
advanced lattice geometries, such as superlattices. In the future, we expect it
to give access to the local full counting statistics, even spin resolved.
Combining such spin-resolved detection with local manipulation of the quantum
gas~\cite{Weitenberg2011} will enable a new generation of experiments with
fermionic quantum matter that can range from the study of multipoint
correlation functions~\cite{Endres2011}, measurement of exotic quasiparticles
and their dynamics~\cite{Cheneau2012} to advanced probing of nonequilibrium
dynamics in many-body systems.

\begin{acknowledgments}
    We thank S.~Blatt and M.~Greiner for helpful discussions and M.~Lohse,
    T.~Gantner, and T.~Reimann for technical assistance while building the
    experimental setup. We acknowledge funding by the MPG and EU (UQUAM).
\end{acknowledgments}

\bibliography{paper}
\section{Supplemental material}

\subsection{Atomic sample preparation}

Our experiments started in a standard magneto-optical trap (MOT) of \Li at
$671\,$nm in a steel octagon vacuum chamber, which was loaded from a Zeeman
slower in $4\,$s. We then illuminated the cloud with MOT beams at $323\,$nm,
driving the $2S_{1/2}\to3P_{3/2}$ transition, which has an effective linewidth
of $2\pi\times750\,$kHz, $8$ times lower than the principal transition at
$671\,$nm and a correspondingly lower Doppler
temperature~\cite{Duarte2011,Sebastian2014}. The ultraviolet (UV) MOT requires
two frequencies, one for cooling and one for repumping out of the $F=1/2$
ground state. After a $12\,$ms UV cooling stage, the atoms reached a
temperature of $70\,\mu$K, which facilitates direct optical trap loading.
During the UV MOT, we ramped up a large volume single beam optical
``collection'' dipole trap, derived from a high power, broadband $1070\,$nm
laser. The collection trap had a power of $65\,$W and a waist of $100\,\mu$m at
the position of the atoms and a wavelength close to the magic one of the UV
transition, where the differential light shift vanishes~\cite{Safronova2012}.
We extinguished the UV repumping beam before the cooling beam to optically pump
the atoms into an incoherent spin mixture of the $F=1/2$ sublevels. We
evaporatively cooled the atoms in the collection trap at a background magnetic
field of $320\,$G, which sets a scattering length of $-290\,\aB$ between the
two spin components. The magnetic field had a curvature confining the atoms
along the axial direction of the trap.  At the end of the $5\,$s linear
evaporation ramp, we transferred the atoms into a tightly focused optical
``transport'' trap at 1064nm ($P=3.5\,$W, $w_0=30\,\mu$m). Its focus was
shifted by movable optics on an air bearing translation stage over $28\,$cm in
$0.5\,$s, transporting the atoms into a glass cell for better optical access.
After this transport, the typically $10^6$ atoms had a temperature of
$13\,\mu$K.

We then ramped up a magnetic field of $320\,$G and a vertical optical ``cross''
trap of $3\,$W power and $w_0=110\,\mu$m that intersected the transport trap
under an angle of $100^\circ$, and applied forced evaporation in this crossed
dipole trap for $3\,$s. A strongly elliptical dimple trap at $780\,$nm with
$w_x = 10.3\,\mu{\rm m}$ and $w_z = 1.7\,\mu{\rm m}$ captured up to $10^4$
atoms from the crossed trap and confined them vertically. Subsequently, we
ramped up the vertical lattice beams, which are derived from a single frequency
$1064\,$nm Yb fiber amplifier. They were focused onto the atoms with a waist of
$w_0=150\,\mu$m and intersected under an angle of $40^\circ$ to generate a
standing wave of $3\,\mu$m spacing and distribute the atoms mainly along a
single plane. As the confinement along the lattice beam direction was very
weak, the vertical cross trap was kept on during this time to give the atoms
additional radial confinement in the plane. Since single plane loading from the
dimple trap did not work with unity fidelity, we need to remove atoms from
adjacent planes. We applied a vertical magnetic field gradient of $18\,$G/cm at
a bias field of $14.6\,$G and used radio frequency transitions to locally
transfer the atoms from $\ket{1/2,\pm1/2}$ to the $\ket{3/2, \mp 1/2}$ states.
These are unstable with respect to spin changing collisions and we confirmed
their loss from the spin flipped planes by radio frequency spectroscopy.

For final evaporative cooling, we ramped up a magnetic bias field of $320\,$G
pointing along a direction in the plane and slowly increased a magnetic field
gradient along the same direction up to $15\,$G/cm in $2\,$s.  The two spin
components experience almost the same magnetic force as their magnetic moment
is nearly equal at this offset field. Therefore, we maintained a balanced and
thermalized spin mixture at the end of the final evaporation.

The horizontal physics lattice was generated by pairs of parallel beams derived
from a $1064\,$nm Nd:YAG laser. They were focused through the high resolution
microscope objective onto the atoms to waists of $180\,\mu$m.  All beam pairs
of the 3D physics lattice had different frequencies to avoid cross-interference
between different lattice axes. The lattice beams were slowly ramped
on after the last evaporative cooling stage, setting the starting point of all
further experiments.

\subsection{Imaging via resolved Raman sideband cooling}

\renewcommand\thefigure{S1}
\begin{figure}[t] \centering
\includegraphics[width=5cm]{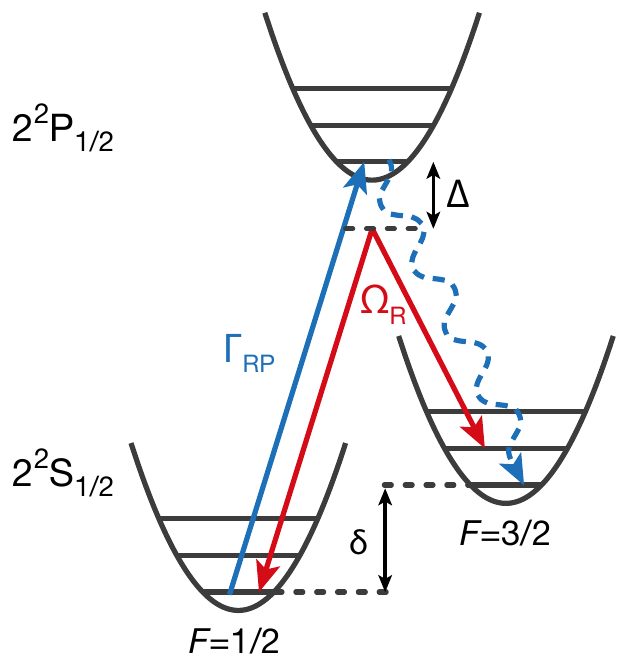}
 \caption{\textbf{Raman sideband cooling scheme.} Two Raman beams, each red
 detuned by $\Delta=7.3\,$GHz to the $D_1$ line, provide the coupling
 $\Omega_{\rm R}$ between different vibrational levels
 $\ket{F=3/2,\,\nu}\Leftrightarrow\ket{F=1/2,\,\nu-1}$. A repump beam (blue)
 optically pumps the atoms from the $\ket{F=1/2,\,\nu-1}$ into the
 $\ket{F=3/2,\,\nu-1}$ state with a scattering rate $\Gamma_{\rm RP}$. The
 hyperfine levels are split by $\delta=228.2\,$MHz and the frequency difference
 between the lowest vibrational levels is $1.3\,$MHz.}
\end{figure}

We used resolved Raman sideband cooling~\cite{Monroe1995,Hamann1998,Kerman2000}
in the pinning lattice to image the atoms.  For our configuration, shown in
Fig.~S1, the Lamb-Dicke parameter is $\eta\approx0.23$, sufficient to suppress
the transfer of atoms to different bands during scattering. The atoms were
continuously illuminated with two linearly polarized Raman beams at a relative
angle of $135^\circ$ and a common red-detuning of $\Delta=7.3\,$GHz with
respect to the $D_1$ line. Their relative detuning was set to couple the
$\ket{F=3/2, \nu}$ and $\ket{F=1/2, \nu-1}$ states, where $\nu$ denotes the
vibrational quantum number. The magnetic field was small (below $10\,$mG),
hence all underlying hyperfine sublevels were degenerate. The Raman beam
orientation was chosen such that the total Raman momentum transfer
$\Delta\vec{k}$ had an equal projection on all pinning lattice axes. This
provides an equal effective coupling strength of $\Omega_{\rm R}\approx 2\pi
\times 90\,{\rm kHz}$ on the red sidebands.  To expand the spatial cooling
region and to couple higher onsite vibrational levels in the anharmonic regime,
we also modulated the two-photon detuning sinusoidally to address on-site trap
frequencies between $900\,{\rm kHz}$ and $1.3\,{\rm MHz}$. The circularly
polarized repump light entered orthogonally to the first Raman beam in the same
plane and was $3\,\Gamma$ blue detuned from the bare atomic transition
$\ket{2S_{1/2}, F=1/2} \to \ket{2P_{1/2}, F=3/2}$, where $\Gamma=2\pi\times
5.8\,$MHz is the natural width of the upper state. This detuning compensated
for the differential light shift in the pinning lattice. In this configuration,
the atoms were optically pumped into the $\ket{F=3/2, \nu=0}$ state, which only
couples to the Raman and repump light via off-resonant scattering.

\subsection{Reconstruction in regions of high filling}

\renewcommand\thefigure{S2}
\begin{figure}[t] \centering
\includegraphics[width=0.5\textwidth]{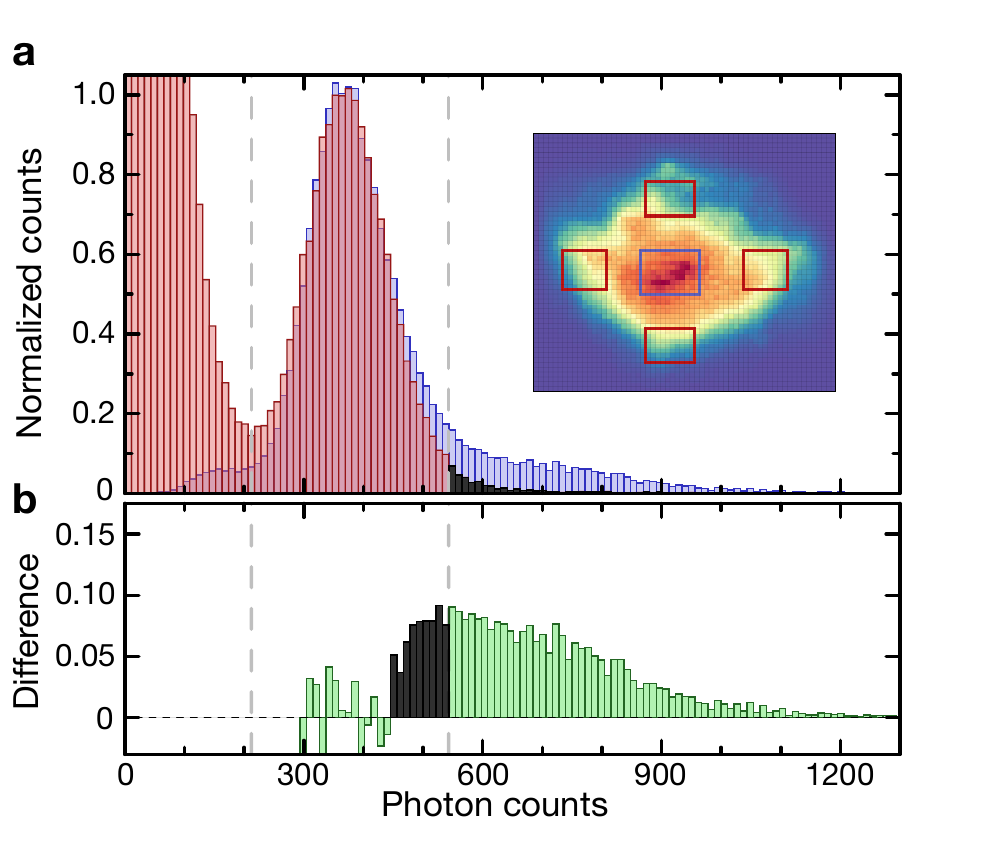}
 \caption{\textbf{Comparison of reconstructed photon counts in regions of high
 and moderate density.} The dashed lines correspond to the thresholds
 discriminating the different occupation numbers. \textbf{(a)} Histograms of
 the amplitude of the local Gaussian PSF fits for the center of the cloud
 (blue) and the edges of cloud (red) normalized to match around and to the left
 of the $n=1$ peak. The black region indicates wrongly assigned double
 occupancies. The regions used to extract the histogram are indicated in the
 inset. Compared to the histogram shown in Fig.~2 of the main text, the signal
 from empty sites in the dense center (left part of the blue histogram) shifts
 to the right. Also, the separation of the zero and one peaks in the red
 moderate filling histogram is slightly reduced. This is due to inhomogeneities
 in the photon count rate across the cloud of approximately $10\%$ and due to
 the increased hopping and loss around the cloud edges. \textbf{(b)} Difference
 between both histograms in the region of high counts. The black area marks
 wrongly assigned single occupancies.}
\end{figure}

In regions of low and moderate filling (up to approximately unity density per
site), the discrimination between zero and one atom per site can be done with
$99\%$ fidelity (cf. main text Fig.~2). The situation becomes more challenging
for higher filling as can be seen in Fig.~S2, where we compare histograms
extracted from different local regions. Instead of a clear peak for $n=0$, the
high density histogram shows a tail at low counts. The leakage of the signal
due to empty sites into the $n=1$ region, determined after subtraction of the
Gaussian fitted single occupation peak, results in an underestimation of the
holes by up to $20\%$.

The main challenge lies in the discrimination between occupations of $n=1$ and
$n=2$. The high density histogram shown in Fig.~S2 displays a long tail to the
right instead of a well defined maximum around twice the single atom count
rate. This causes difficulties when determining an accurate threshold between
$n=1$ and $n=2$. We set this threshold such that the mean count rate of single
atoms is in the center of the $n=1$ region. While there is a significant amount
of signal well above this threshold, there are many occurrences around the
threshold value. This is the main source for systematic errors in determining
the correct occupation number for these photon counts.

To estimate these errors, we compare the histogram extracted from the central
high density region to the ones of moderate density regions along the edge of
the cloud. We find a clear difference for higher photon counts between them
(see Fig.~S2). For the comparison, we normalized the histograms to each other,
such that they overlap around the center and left slope of the $n=1$ peak.
Then, we subtracted both histograms to get a residual signal that we attribute
to the double occupancies. Of these residual counts, the events below the
threshold are interpreted as double occupancies that are falsely assigned as
single atoms. This region has an area which is around $30\,\%$ of the area
above the threshold, therefore our measured number of double occupancies is
underestimated by up to this percentage. We also analyzed the signal from the
single atoms at the edge of the system that resulted in a few events exceeding
the threshold. These constitute around $2.5\,\%$ of the total single atom
signal and represent single atoms falsely identified as double occupancies.

Furthermore, each wrong assignment of a doubly occupied site results in the
opposite wrong assignment of a single occupancy. We took these anti-correlated
systematic errors in the probabilities for single and double occupancies into
account when analyzing the uncertainty of the local mean, number fluctuations
and entropy. These systematic uncertainties are given by the shaded areas in
our plots in the main text.

\subsection{Entropy analysis}

Given the measurement of the local occupation probabilities, we extracted the
entropy per site $S$ (we drop the site index for better readability). Our
analysis is based on the many-body states (with occupation probabilities $p_E$)
of up to two fermions on $k$ single particle levels with occupation
probabilities $y_i$. The corresponding entropy per site is then given by
$S/k_{\rm B}=\sum_E p_E \mathrm{ln}p_E$. In the main text we present the
results for $k=2$, for which the population of the local energy eigenstates are
\begin{eqnarray}
  p_0 &=& (1-y_0)(1-y_1) = x_0 \nonumber \\
  p_1 &=& y_0 (1-y_1)\nonumber \\
  p_2 &=& y_1 (1-y_0)\nonumber \\
  p_3 &=& y_0 y_1 = x_2 \,. \nonumber
\end{eqnarray}

Here, $p_0$ and $p_3$ are the probabilities for zero and two fermions,
corresponding to the lowest and highest energy eigenstate. The probabilities
for the remaining two eigenstates are $p_1$ and $p_2$ and describe a single
fermion in the lower and upper single particle level, respectively.  All $p_E$
can be calculated from the measured probabilities for holes and double
occupancies, $x_0$ and $x_2$ respectively.\\ We extended the analysis in the
same spirit to the case $k=4$, assuming an equal population of the three upper
single particle levels. These upper levels correspond to the first excited
states of the local three dimensional harmonic oscillator.  This gives a
reasonable upper bound for the inferred entropy. In regions of low double
occupancy, this has a minor effect on the detected entropy, leading to an
increase from $0.3\,k_{\rm B}$ to $0.34\,k_{\rm B}$ of the minimal measured
entropy per particle. In the center of the cloud, where the fraction of double
occupancies is higher, the inferred entropy increases from $0.6\,k_{\rm B}$ to
$0.95\,k_{\rm B}$.

\end{document}